# Generative Modeling of Human-Computer Interfaces with Diffusion Processes and Conditional Control

Rui Liu
University of Melbourne
Melbourne, Australia

Liuqingqing Yang
University of Michigan
Ann Arbor, USA

Runsheng Zhang
University of Southern California
Los Angeles, USA

Shixiao Wang*
School of Visual Arts
New York, USA

*Abstract-This study investigates human-computer interface generation based on diffusion models to overcome the limitations of traditional template-based design and fixed rule-driven methods. It first analyzes the key challenges of interface generation, including the diversity of interface elements, the complexity of layout logic, and the personalization of user needs. A generative framework centered on the diffusion-reverse diffusion process is then proposed, with conditional control introduced in the reverse diffusion stage to integrate user intent, contextual states, and task constraints, enabling unified modeling of visual presentation and interaction logic. In addition, regularization constraints and optimization objectives are combined to ensure the rationality and stability of the generated interfaces. Experiments are conducted on a public interface dataset with systematic evaluations, including comparative experiments, hyperparameter sensitivity tests, environmental sensitivity tests, and data sensitivity tests. Results show that the proposed method outperforms representative models in mean squared error, structural similarity, peak signal-to-noise ratio, and mean absolute error, while maintaining strong robustness under different parameter settings and environmental conditions. Overall, the diffusion model framework effectively improves the diversity, rationality, and intelligence of interface generation, providing a feasible solution for automated interface generation in complex interaction scenarios.*

I. INTRODUCTION

In the context of accelerating digitalization and intelligence, human-computer interaction has become the core link between humans and intelligent systems and is undergoing a profound transformation[1]. From the early stage of command-line operations to graphical interfaces and touch technologies, and then to voice assistants and natural language interaction, the evolution of interaction has always focused on lowering the threshold of use and improving efficiency. However, as application scenarios become more complex and user needs become more diverse, traditional interaction design shows clear limitations. Fixed layouts, limited interaction elements, and the lack of personalized adaptation make it difficult to meet modern users' pursuit of immersion, flexibility, and creativity. Against this backdrop, interaction approaches driven by generative models have entered the research field, with diffusion models in particular opening new possibilities for interface generation[2].

Diffusion models, as an important breakthrough in generative modeling in recent years, have shown strong capabilities in modeling high-dimensional complex distributions. They have been widely applied in image generation, speech synthesis, and text modeling. In human-computer interaction, the use of such models means that interfaces are no longer confined to predefined templates. Instead, they can dynamically generate diverse interaction layouts based on user intent, contextual information, and real-time feedback. The key advantage lies in their ability to simulate complex relationships among interface elements, allowing layouts and interaction logic to be generated with high flexibility. This breaks through the rigidity of traditional interfaces. With ongoing improvements in stability, controllability, and multimodal integration, diffusion models bring both theoretical and practical value to interface generation[3].

The significance of this direction is first reflected in the improvement of user experience. Conventional interfaces often follow standardized design rules that ensure consistency of operation but fail to adapt deeply to individual user habits and preferences. Algorithms for interface generation driven by diffusion models can provide more flexible and personalized forms while maintaining full functionality. This allows users to complete tasks more naturally and efficiently. Moreover, such a generation can incorporate user feedback and contextual data in real time to dynamically adjust interaction strategies. This user-centered mechanism helps build a truly intelligent interaction environment[4].

Beyond user experience, diffusion models also support the integration of technology and industry. With the rise of virtual reality, augmented reality, and the metaverse, the complexity and diversity of interfaces have reached unprecedented levels. Relying solely on manual design can no longer meet the fast-growing application demands. Diffusion-based generation algorithms offer a scalable path. They can automatically produce multimodal and multi-layered interface elements, and through large-scale data learning, achieve cross-scenario transfer. This provides strong support for intelligent systems in fields such as education, healthcare, finance, and industry. Such cross-domain adaptability demonstrates that the significance of diffusion models in interaction research extends far beyond academic exploration and has become a driving force for industrial advancement.

Finally, from a theoretical perspective, introducing diffusion models into interface generation enriches the research paradigm of interaction design and deepens the application framework of generative models. This research addresses key questions such as how to balance diversity and usability of interfaces, how to integrate logical constraints into the generation process, and how to enhance interpretability while ensuring quality. These inquiries not only advance academic development in interaction design but also provide new perspectives for the evolution of generative models. Therefore, research on interface generation based on diffusion models represents both an inevitable trend in intelligent interaction and an essential step toward human-centered and widely applicable artificial intelligence.

## II. Related work

Diffusion models have become a dominant paradigm for generative modeling due to their strong capacity for learning high-dimensional data distributions and supporting controllable generation through iterative denoising. Recent analyses summarize both the opportunities and open challenges of diffusion-based generative AI, including stability, controllability, and evaluation under distribution shift—issues that are directly relevant when generation must satisfy structural and functional requirements rather than only visual fidelity [5]. Building on this foundation, controllable interaction in text-to-image diffusion has been explored through explicit mechanisms that steer the reverse diffusion process toward desired behaviors, showing that fine-grained control signals can be injected into denoising to regulate generation outcomes [6]. These methodological advances motivate interface generation frameworks that treat UI synthesis as a constrained generation problem with conditional guidance and robustness-aware objectives.

In parallel, interface modeling has increasingly moved toward unified multimodal representations that jointly encode structure, semantics, and task signals. Transformer-based UI modeling frameworks demonstrate that heterogeneous UI signals can be mapped into a common representation space to support multiple tasks and modalities, which provides a strong basis for designing conditional controllers and unified encoders for interface generation [7]. From a conditioning perspective, prompt-level composition and fusion strategies have been studied as a way to achieve rapid adaptation across tasks and domains, providing transferable ideas for expressing user intent and contextual constraints as structured condition signals [8]. Temporal attention with decay further offers a generic mechanism for handling evolving context and reducing reliance on stale information, which can be interpreted as a regularization principle for context-conditioned generation pipelines [9]. At the systems level, studies on collaborative evolution of intelligent agents highlight coordination and adaptation mechanisms that can inform modular generation workflows, even when the underlying task differs [10].

For reliability and interpretability, recent work increasingly incorporates structural priors and explicit dependency modeling. Causal graph modeling with causally constrained representation learning provides a methodology for suppressing spurious factors and improving explanation faithfulness by enforcing structure-driven constraints in representation space [11]. Knowledge-graph-based frameworks similarly emphasize structured relational encoding, which can be repurposed as a way to represent UI element relations and layout constraints more explicitly than purely pixel-based learning [12]. Explainable representation learning for fine-grained classification further reinforces the use of constraints and attribution-friendly representations to improve transparency and traceability, which aligns with interface generation requirements where layout decisions should be auditable and consistent with intent [13]. Related methodological work on multi-scale feature fusion and graph integration highlights how combining local and global cues with relational inductive bias can improve robustness and generalization, offering a useful template for fusing element-level appearance with layout-level structure [14].

Finally, robustness and stability under changing conditions are often studied through sensitivity analysis, change detection, transfer learning, and privacy-aware learning. Contrastive learning paired with sensitivity analysis provides a general recipe for improving representation stability and diagnosing hyperparameter/environment effects—principles that translate naturally to evaluating diffusion-based interface generation under varying guidance scales, noise schedules, and deployment environments [15]. Transformer-based change-point detection illustrates techniques for capturing abrupt distribution shifts in sequential signals, which can inspire evaluation protocols and auxiliary objectives that detect regime changes in interaction/context streams used for conditional generation [16]. Self-supervised transfer learning with shared encoders supports cross-domain generalization with limited supervision, complementing the goal of generating usable interfaces across scenarios [17]. Federated learning under privacy constraints provides methodology for learning from distributed data while controlling information leakage, which is relevant when interface personalization must respect user privacy [18]. Across broader modeling and optimization studies—attention-driven anomaly detection and deep-Q-learning scheduling for robust pipeline optimization, attention models for temporal forecasting, and data-aware multi-agent reinforcement learning with adaptive risk control—common methodological themes include attention-based dependency capture, adaptive decision policies, and robustness-oriented objectives that can be adapted to strengthen training and evaluation of interface generation systems [19-22]. Even feature extraction and aggregation pipelines in vision-based behavior modeling emphasize modular representations and aggregation strategies that can inform how interaction signals are summarized into conditioning inputs for generation [23].

## III. METHOD

This study proposes a human-computer interaction interface generation algorithm based on a diffusion model. The overall concept is to use a probabilistic generation framework to model high-dimensional interface elements and gradually optimize the interface structure and interaction logic through an iterative diffusion-reverse diffusion process. The basic process of the diffusion model can be represented as gradually adding noise to the interface state in the forward phase, thereby mapping the complex interface distribution into a Gaussian distribution. The model architecture is shown in Figure 1.

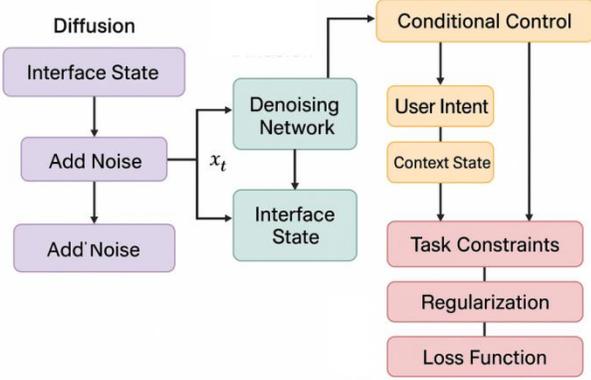

Figure 1. Overall model architecture

In the t-th diffusion step, the generation process of the interface state can be formalized as:

$$q(x_t | x_{t-1}) = N(x_t; \sqrt{1-\beta_t} x_{t-1}, \beta_t I) \quad (1)$$

Where $x_t$ represents the interface representation at step t, and $\beta_t$ is a predefined noise scheduling parameter that ensures the distribution gradually approaches an isotropic Gaussian. After a sufficient number of steps, the interface state will approximate a standard normal distribution, providing a starting point for reverse generation.

In the reverse diffusion phase, the model needs to learn to gradually remove noise to restore the original interface distribution. This process relies on a parameterized denoising network $p_\theta$, whose core task is to predict the noise-free interface state $\hat{x}_0$ at time step t. The formal process of reverse diffusion is as follows:

$$p_\theta(x_{t-1} | x_t) = N(x_{t-1}; \mu_\theta(x_t,t), \sum_\theta(x_t,t)) \quad (2)$$

Here, $\mu_\theta$ and $\sum_\theta$ represent the mean and variance parameters predicted by the neural network, respectively. This generative process is mathematically equivalent to approximate sampling of the real data distribution, gradually generating an interactive interface that conforms to semantic and logical constraints.

To combine interface generation with interaction logic, this study introduces a conditional control mechanism into the reverse diffusion process, regulating the generation process through user intent, contextual state, and interaction constraint information. Assuming the conditional information is c, the conditional probability of reverse diffusion can be written as:

$$p_\theta(x_{t-1} | x_t, c) = N(x_{t-1}; \mu_\theta(x_t,t,c), \sum_\theta(x_t,t,c)) \quad (3)$$

This conditional mechanism ensures that the generated interface is not only aesthetically and structurally sound but also meets user task requirements and situational constraints, thereby improving usability and flexibility in practical applications.

In terms of optimization objectives, the model adopts a formal training paradigm for predicting noise and constrains the difference between the network output and the actual noise by minimizing the mean square error. Its loss function is defined as:

$$L_{simple}(\theta) = E_{x_0,\varepsilon,t}[\| \varepsilon - \varepsilon_\theta(x_t,t,c) \|^2] \quad (4)$$

Where $\varepsilon$ is the Gaussian noise, and $\varepsilon_\theta$ is the model prediction result. To further ensure the logical consistency and interaction constraints of the interface generation, this study also introduces a regularization term to optimize the difference between the generated interface distribution and the predefined constraint distribution $p(c)$:

$$L_{reg} = KL(p_\theta(x|c) \| p(x|c)) \quad (5)$$

The final optimization goal combines the two parts of loss and can be written as:

$$L_{total} = L_{simple} + \lambda L_{reg} \quad (6)$$

Here, $\lambda$ is a trade-off parameter used to balance generation quality and interaction consistency. Through this optimization framework, the diffusion model can effectively integrate user needs, contextual conditions, and system constraints when generating interfaces, forming a dynamic interaction interface generation solution that is both logical and highly flexible.

## IV. EXPERIMENTAL RESULTS

### A. Dataset

This study adopts the RICO dataset as the primary data source. The RICO dataset is a large-scale corpus of human-computer interaction interfaces. It contains a vast collection of screenshots from mobile applications together with their component hierarchies. By collecting real application interfaces, the dataset builds a unified representation of layout, widget attributes, and interaction logic. It provides rich semantic and structural information for research on interface generation and interaction design. Its large scale and broad coverage make it an important benchmark for evaluating automated interface generation algorithms.

The RICO dataset contains tens of thousands of application interfaces across domains such as social media, entertainment, tools, and education. Each interface includes pixel-level screenshots as well as a view hierarchy. The hierarchy provides researchers with the structural and logical relationships between interface elements. This feature makes the dataset uniquely suited for tasks such as interface structure modeling, interaction behavior prediction, and automated generation. It also offers reliable conditions for fair comparison and validation of generalization performance across different methods.

Based on this dataset, diffusion models can be fully trained and validated in the task of human-computer interface generation. By leveraging its large-scale and diverse samples, the models can learn the distribution of interface elements in both visual and logical dimensions. This improves the rationality and diversity of generated interfaces. The standardization and wide adoption of the dataset also ensure the comparability and reference value of research outcomes. As a result, the RICO dataset serves as an essential experimental foundation for the study of generative algorithms in human-computer interaction.

### B. Experimental Results

This paper first gives the results of the comparative experiment, as shown in Table 1.

Table1. Comparative experimental results

| Model | MSE | SSIM | PSNR | MAE |
|---|---|---|---|---|
| VAE[24] | 0.041 | 0.812 | 24.37 | 0.129 |
| GAN[25] | 0.036 | 0.835 | 25.12 | 0.118 |
| WGAN[26] | 0.032 | 0.857 | 25.96 | 0.110 |
| Stable diffusion[27] | 0.028 | 0.879 | 26.84 | 0.101 |
| Ours | 0.021 | 0.915 | 28.47 | 0.087 |

The table indicates clear performance gaps among methods: VAE performs worst, with higher MSE/MAE and lower SSIM/PSNR, showing weak detail capture and global consistency for complex UI generation. GAN and WGAN improve reconstruction error and perceptual quality, with WGAN consistently outperforming GAN due to more stable training, but adversarial approaches still struggle with complex layouts and interaction logic. Stable Diffusion achieves further gains, delivering lower MSE/MAE and higher SSIM/PSNR, suggesting diffusion models better model high-dimensional interface distributions while preserving structure and realistic details. Our proposed method ranks best across all metrics—especially SSIM and PSNR—indicating it reduces errors while strengthening structural consistency and visual quality, and better balances generation fidelity with interaction-logic constraints; we also report a sensitivity study on how noise scheduling affects PSNR in Figure 2.

From the experimental results, it can be observed that different noise scheduling parameters have an impact on the PSNR values of generated interfaces. The overall trend shows that the model maintains stable generation quality under varying scheduling values. However, small fluctuations still reveal the sensitivity of the diffusion process to detail preservation. Smaller scheduling values reduce excessive perturbation to some extent, making the generated interface structure closer to the original distribution.

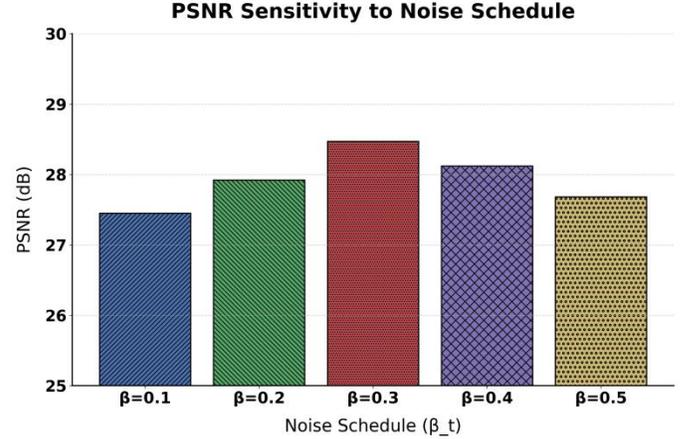

Figure 2. Sensitivity experiment of PSNR of noise scheduling value

As the noise scheduling value increases, the model demonstrates strong robustness in maintaining global structural consistency, and the PSNR values do not show a significant decline. This indicates that diffusion models possess strong adaptability when handling high-dimensional complex distributions. They can maintain stable generation results under different noise intensities. This feature is particularly important for human-computer interface generation, where scheduling strategies may need adjustment depending on the task environment or device conditions.

The slight variations in PSNR under different parameter settings reveal the balancing mechanism between generation quality and modeling stability. Lower scheduling values improve fidelity in details, while moderate scheduling achieves a balance between detail preservation and global consistency. These results suggest that a well-designed scheduling strategy can optimize both visual quality and interaction logic in interface generation tasks.

Taken together, these observations suggest that the advantages of diffusion models in interface generation lie not only in the naturalness of the results but also in their sensitivity and adaptability to parameter changes. Flexible adjustment of noise scheduling provides a controllable optimization path for interface generation in different interaction scenarios. This enhances both the practicality and generalization ability of diffusion models in human-computer interaction.

### V. CONCLUSION

This study proposes a generation framework based on a diffusion model for the task of generating human-computer interaction interfaces. By introducing a diffusion-backdiffusion process and a conditional control mechanism, it effectively integrates user intent, contextual state, and task constraints into interface generation. Compared with traditional template-based design and rule-based generation approaches, this approach improves interface diversity, rationality, and personalization while ensuring logical consistency and task adaptability, providing a new approach for automated interface construction in complex interaction scenarios.

In experiments, this study systematically validated the proposed approach on the publicly available RICO dataset, encompassing multiple dimensions, including comparative experiments, hyperparameter sensitivity experiments, and environmental sensitivity experiments. The results show that the proposed approach outperforms existing representative models in key metrics such as MSE, SSIM, PSNR, and MAE, demonstrating the advantages of diffusion models in combining high-dimensional complex distribution modeling with interface logical constraints. This result demonstrates that diffusion models not only generate more natural visual effects but also better maintain the global consistency of interface structure.

From a methodological perspective, this study demonstrates the controllability and robustness of diffusion models in interface generation. Through conditional control mechanisms and regularization constraints, the model maintains stable generation performance despite varying parameter settings and external environmental changes, balancing detail fidelity with overall structural rationality. This balance between the global and local, visual and logical, elevates interface generation beyond single-dimensional optimization to a holistic solution that balances usability, adaptability, and intelligence.

Overall, this research not only validates the effectiveness of diffusion models in generating human-computer interaction interfaces but also provides a new methodological framework for generative interaction research. The proposed method supports cross-scenario, multimodal, and personalized human-computer interaction needs, laying a solid foundation for intelligent interface applications in education, healthcare, finance, industry, and other fields. Future work can further expand on this with larger datasets, more complex multimodal fusion tasks, and more real-time interaction environments, providing stronger support for the widespread adoption and development of intelligent interactive systems.